\documentclass{aa}  
\usepackage{graphicx}
\usepackage{txfonts}
\usepackage[hidelinks]{hyperref}
\usepackage{amsmath}% Advanced maths commands
\usepackage{amssymb}    % Extra maths symbols
\usepackage[T1]{fontenc}
\usepackage{ae,aecompl}
\usepackage{xcolor}
\usepackage{multirow}
\usepackage{tabularx}
\usepackage{threeparttable}
\usepackage{natbib}
\usepackage{adjustbox}
\bibpunct{(}{)}{;}{a}{}{,} % to follow the A&A style

\usepackage{siunitx}

\begin{document}

   \title{Reflection component in the Bright Atoll Source GX 9+9}

 %  \subtitle{}
% \author{R. Iaria\inst{1} et al.}
 
   \author{R. Iaria\inst{1}, S. M. Mazzola\inst{1}, T. Di Salvo\inst{1}, A. Marino\inst{1,3,4}, A.F. Gambino\inst{2},  A. Sanna\inst{2}, A. Riggio\inst{2},  \\  L. Burderi\inst{2}}

  \institute{Dipartimento di Fisica e Chimica - Emilio Segrè, Universit\`a di Palermo, via Archirafi 36 - 90123 Palermo, Italy
        \and
         Dipartimento di Fisica, Universit\`a degli Studi di Cagliari, SP Monserrato-Sestu, KM 0.7, Monserrato, 09042 Italy
  \and
  Istituto Nazionale di Astrofisica, IASF Palermo, Via U. La Malfa 153, I-90146 Palermo, Italy
  \and
  IRAP, Universitè de Toulouse, CNRS, UPS, CNES, Toulouse, France  }
   
% \abstract{}{}{}{}{} 
% 5 {} token are mandatory
 
  \abstract
  % context heading (optional)
  % {} leave it empty if necessary  
   {GX 9+9 (4U 1728-16) is a low mass X-ray binary (LMXB) source harboring a neutron star. Although it belongs to the subclass of the bright Atoll sources together with GX 9+1, GX 3+1, and GX 13+1, its broadband spectrum is poorly studied and apparently does not show reflection features in the spectrum. }
  % aims heading (mandatory)
   {To  constrain the continuum well and verify whether a relativistic smeared reflection component is present, we analyze the broadband spectrum of GX 9+9 using {\it BeppoSAX} and \textit{XMM-Newton} spectra covering the 0.3-40 keV energy band.}
  % methods heading (mandatory)
   {We fit the spectrum adopting a model composed of a disk-blackbody plus a Comptonized component whose seed photons have a blackbody spectrum (Eastern Model). A statistically equivalent model is composed of a Comptonized component whose seed photons have a disk-blackbody distribution plus a blackbody that mimics a saturated Comptonization likely associated with a boundary layer (Western model). Other trials did not return a good fit.}
  % results heading (mandatory)
   { The spectrum of GX 9+9 was observed in a soft state and its luminosity is $2.3 \times 10^{37}$ erg s$^{-1}$ assuming a distance to the source of 5 kpc. In  the  Eastern Model scenario, we find the seed-photon temperature and electron temperature of the Comptonized component to be $1.14^{+0.10}_{-0.07}$ keV and $2.80^{+0.09}_{-0.04}$ keV, respectively, while the optical depth of the Comptonizing corona is $8.9\pm0.4$. The color temperature of the inner accretion disk is $0.86^{+0.08}_{-0.02}$ keV and $0.82 \pm 0.02$ keV for the {\it BeppoSAX} and \textit{XMM-Newton} spectrum, respectively. In  the  Western Model scenario, instead, we find that the seed-photon temperature is $0.87 \pm 0.07$ keV and $1.01 \pm 0.08$ keV for the {\it BeppoSAX} and \textit{XMM-Newton}  spectrum, respectively. The electron temperature of the Comptonized component is $2.9\pm0.2$ keV,   while the optical depth is $9.4^{+1.5}_{-1.1}$. The blackbody temperature is $1.79^{+0.09}_{-0.18}$ keV and $1.85^{+0.07}_{-0.15}$ keV for the {\it BeppoSAX} and \textit{XMM-Newton}  spectrum, respectively. The addition of a relativistic smeared reflection component improved the fit in both the scenarios, giving compatible values of the parameters, even though a significant broad emission line in the Fe-K region is not observed.}
  % conclusions heading (optional), leave it empty if necessary 
  { From the reflection component we estimated an inclination angle of about $43^{+6}_{-4}$ deg and $51^{+9}_{-2}$ deg for the Eastern and Western Model, respectively. The value of the reflection fraction $\Omega/2\pi$ is $0.18\pm0.04$ and $0.21\pm0.03$ for the Eastern and Western Model, respectively, suggesting that the Comptonized corona should be compact and close to the innermost region of the system.}

  \authorrunning{R. Iaria et al.}

  \titlerunning{Reflection component in the Bright Atoll Source GX 9+9}
  
  
 %\object{GX 9+9}

   \maketitle

\section{Introduction}
A number of bright X-ray emitting neutron stars (NSs) in accreting
binary systems is now widely known and is being studied \citep[see][]{Liu_01}. Their X-ray
emission is given by accretion of matter transferred by the
companion star on the NS surface, and often via an accretion
disk. Generally, the spectral properties of such sources are complex and usually characterized by a multi-color 
blackbody component and a high energy Comptonization component. The
former is likely associated with the emission of an accretion disk close to the NS, while the latter can be ascribed to the contribution of the NS surface and/or a corona, with an emitting geometry that is  relatively poorly understood. However, a line of evidence suggests that it is produced in the inner regions of the system, possibly around the NS \citep[as in a boundary layer, e.g.,][]{dai_10,Pintore_15}. 
Nevertheless,  an alternative model of the continuum is possible in which the seed photons for Comptonization originate from an accretion disk and the blackbody component is associated with a hot boundary layer/NS surface emission
\citep[Western model; e.g.,][]{White_88}.

The spectra of some bright X-ray sources harboring a NS
also show a hard power-law component dominant above 20 keV \citep[e.g.,][]{Disalvo_01,Iaria_01,Paizis_06,Piraino_07,Pintore_14,Pintore_16,Reig_16} that changes its intensity in time until it  disappears \citep[see][]{Iaria_04}.
The origins of this component are still matter of debate. It has been suggested that such features can be produced by either bulk motion of matter around the NS \citep{Titarchuk_98}, or
by Compton scattering in hybrid thermal and/or non-thermal medium
\citep{Putanen_98} or by  synchrotron self-Compton emission at the base
of an outflowing jet \citep{Markoff_01}. The first scenario was adopted by \cite{Mainardi_10} to model the spectra of the X-ray bright sources GX 5-1, GX 13+1, GX 3+1, GX 9+1, and GX 349+2, while the second scenario was adopted to describe the spectral evolution of Sco X-1 along its color-color diagram \citep{Dai_07}. \cite{Mainardi_10} adopted two Comptonized components to model the spectra of these sources: the first  is associated with the inner region of the accretion disk that is inflated because of the high accretion rate; instead, the second  is related to a transient layer going from the NS to the innermost region of the disk, where the matter could have a bulk motion falling onto the NS. The authors found that the first Comptonization component describes the dominant part of the spectrum, and  is interpreted as thermal Comptonization of soft seed photons (<1 keV), likely from the accretion disk, by a 3–5 keV corona. Instead, the second Comptonization component varies dramatically, spanning from bulk plus thermal Comptonization of blackbody  seed photons to a blackbody emission. 

In addition to the above-mentioned continuum, discrete emission
features are often observed in the high quality spectra of NS bright X-ray sources. These features are usually broad (width <1 keV), and the most common is associated with  the Fe K$\alpha$ line transitions at 6.4–6.9 keV. They are produced by photoelectric absorption and subsequent fluorescent emission of the hard photons coming from the Comptonized corona off the neutral or ionized matter in the top layers of the accretion disk  \citep[see, e.g.,][and references therein]{Fabian_00,Piraino_12,Chiang_16,Iaria_16,Matranga_17,Mazzola_19}. The broadening is due to the fast motion of matter in the innermost region of the disk which causes relativistic Doppler and gravitational redshift. These relativistic features are also accompanied by a reflection continuum characterized by a Compton hump around 20-40 keV. It can be ascribed to direct Compton scattering, and it is observed mainly when the source is in a hard spectral state because the flux above 20 keV is higher with respect to the soft state \citep[see][]{Miller_13,Egron_13,Iaria_19}. 

We  expect the presence of a reflection component in all
the galactic X-ray bright sources, but surprisingly this is not the case.
For example,  \citet{Homan_18} did not find
a signature of a  reflection spectral component for the source GX 5-1; 
they note that a broad Fe line was also observed in the {\it XMM-Newton} spectra of the Z source GX 340+0 \citep{Dai_09}, which is considered a ``twin'' of GX 5-1.  The authors suggest, as more plausible explanation of this peculiar behavior, that for an ionization parameter $\xi$ larger than $10^4$, as deduced for GX 5-1, the Fe K$\alpha$ line becomes weak and broadened by Compton scattering and then particularly difficult to detect. 

\section{Source GX 9+9} 
GX 9+9, also known as 4U 1728-16, is a low mass X-ray binary (LMXB) discovered during a sounding rocket observation of the Galactic center \citep{Bradt_68}. Together with GX 9+1, GX 3+1, and GX 13+1, the source has been classified as a bright atoll-type neutron star binary \citep{Hasinger_89}.
The system consists of a NS accreting mass from a low main sequence star identified as an early M-class dwarf with mass and radius of 0.2–0.45 M$_{\odot}$ and 0.3–0.6 R$_{\odot}$, respectively \citep{Hertz_88,Schaefer_90}. An orbital period of $4.19 \pm 0.02$ hr was discovered by \cite{Hertz_88}, analyzing data taken with the High Energy Astronomy Observatory-1, and by \cite{Schaefer_90} in the optical band. \cite{Hertz_88}, assuming a NS mass of 1.4 M$_{\odot}$, used the lack of eclipses in the X-ray light curve to show that the inclination angle of GX 9+9 is less than $63^{\circ}$. 
The distance to the source is not well established yet; however, since the system is considered to be a Galactic bulge object, a distance of 5-7 kpc is usually adopted \citep{vilhu_07}.

Although GX 9+9 is a bright source, its spectrum is poorly studied. \cite{Church_01}, using ASCA data, fitted the 0.7-10 keV spectrum adopting a cutoff power-law component with a cutoff energy at 4.1 keV and absorbed by an equivalent hydrogen column of neutral interstellar matter of $2.1 \times 10^{21}$ cm$^{-2}$. The authors estimated a total 1-30 keV luminosity close to $2.5 \times 10^{37}$ erg s$^{-1}$, assuming a distance to the source of 5 kpc. 
\cite{Kong_06} investigated the continuum of the source
by modeling the 2.5-20 keV RXTE spectra (observed with the PCA instrument between 11 and 14 August 1999) with a combination of a
Comptonized component and a blackbody, both of which were absorbed by neutral matter having an equivalent hydrogen column kept fixed to the value 2.1$\times 10^{21}$ cm$^{-2}$ reported by \citeauthor{Church_01}.
Analysing 24 spectra, the authors found a blackbody temperature between 1.4 and 2.0 keV, an electron temperature between 2.8 keV and 3.4 keV, and an optical depth between 9 and 12. The 2-20 keV flux was between $6.6 \times 10^{-9}$ and $8.7 \times 10^{-9}$ erg s$^{-1}$ cm$^{-2}$ (corresponding to a luminosity between $2.0 \times 10^{37}$ and $2.6 \times 10^{37}$ erg s$^{-1}$ for a distance to the source of 5 kpc).
\begin{figure}
\centering
\includegraphics[scale=.55]{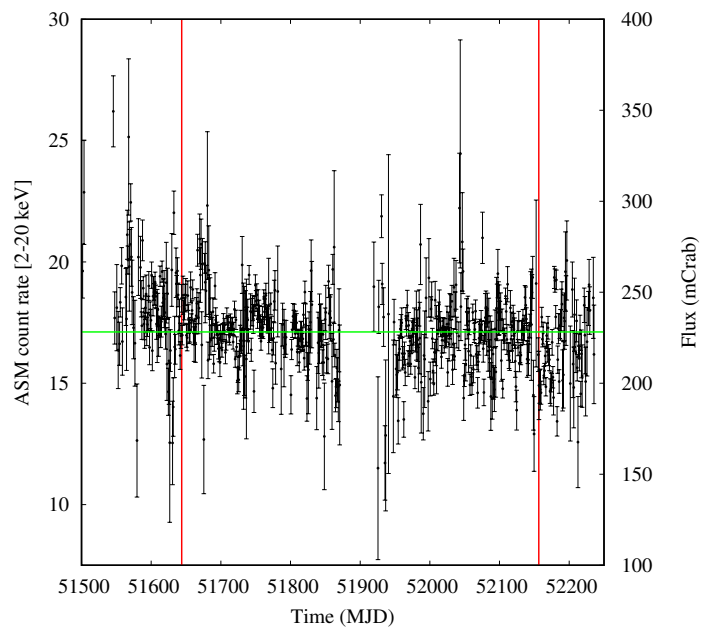}
\caption{ASM light curve of GX 9+9 between 51500 and 52250 MJD. The bin time is one day. The two vertical red lines indicate the times of the \textit{BeppoSAX} and \textit{XMM-Newton} observations, respectively. The horizontal green line indicates the mean ASM count rate of 17.1 c s$^{-1}$ (228 mCrab).}
\label{figure:ASMlc}
\end{figure}

\cite{Gogus_07} analyzed the RXTE spectrum of GX 9+9 (collected on 16 October 1996) in the 2-20 keV energy range adopting a model composed of two blackbody components with temperatures of $2.02 \pm 0.02$ keV and $0.88 \pm 0.02$ keV, respectively. The 2-20 keV unabsorbed flux was roughly $4.7 \times 10^{-9}$ erg s$^{-1}$ cm$^{-2}$, corresponding to a luminosity {\bf of} $1.4 \times 10^{37}$ erg s$^{-1}$ for a distance to the source of 5 kpc. 

\cite{Savolainen_09}, using 196 spectra taken with the RXTE and INTEGRAL observatory, fitted the 3-40 keV spectrum of GX 9+9 adopting a model composed of thermal emission coming from the inner region of the accretion disk plus a blackbody representing the spreading layer, which is an extended accretion zone on
the NS surface, as opposed to the more traditional ring-like boundary layer. The authors suggested that Comptonization was not required to adequately fit the spectrum and that a distance to the source of 10 kpc could clearly explain the best-fitting normalization of the accretion disk blackbody component. 

In this work we present the first broadband analysis of GX 9+9
in the 0.3-40 keV energy range using \textit{BeppoSAX}and \textit{XMM-Newton} data. We show that the continuum is well fitted using a model composed of a thermal emission from the inner region of the accretion disk plus a Comptonized component with electron temperature at 2.9 keV.
Although   no clear residuals  of a broad emission line in the Fe-K region  are evident, we found that a reflection component is needed to obtain a good fit. 
%%%%%%%%%%%%%%%%%%%%%%%%%%%%%%%%%%%%%%%%%%%%%

\section{Observations and data reduction}
We analyzed the observations of GX 9+9 collected by the \textit{BeppoSAX} satellite and the \textit{XMM-Newton} observatory in 2000 and 2001, respectively. Although the observations are separated by  more than a year, our aim is  to fit simultaneously the spectra 
obtained from the two observations in order to study the broadband spectrum in the 0.3-50 keV energy range. To do that we verified 
that the observed flux and the hardness ratio of the source were similar by analyzing the light curve taken by  All Sky Monitor (ASM) on board the RXTE observatory. 
We found that the observed flux is similar in the two periods, the ASM light curve of GX 9+9 (Fig. \ref{figure:ASMlc}), taken between 51500 and 52250 MJD, shows a mean count rate close to 17.1 c s$^{-1}$ (horizontal green line) corresponding to a flux of 228 mCrab ($\sim 6.5 \times 10^{-9}$ erg cm$^{-2}$ s$^{-1}$). The two vertical red lines indicate the times of the \textit{BeppoSAX} and \textit{XMM-Newton} observations, respectively. 
Furthermore, we verified that the ASM hardness ratios (HRs) $[3-5 {\rm keV}]/[1.3-3 {\rm keV}]$ and  $[5-12 {\rm keV}]/[3-5 {\rm keV}]$ were similar
during the \textit{BeppoSAX}  and   \textit{XMM-Newton} observations. We find a good agreement between the two values of the HRs obtained during the two observations, as we expect for Bright Atoll sources.   
%%%%%%%%%%%%%%%%
\begin{figure}
\centering
\includegraphics[scale=.65]{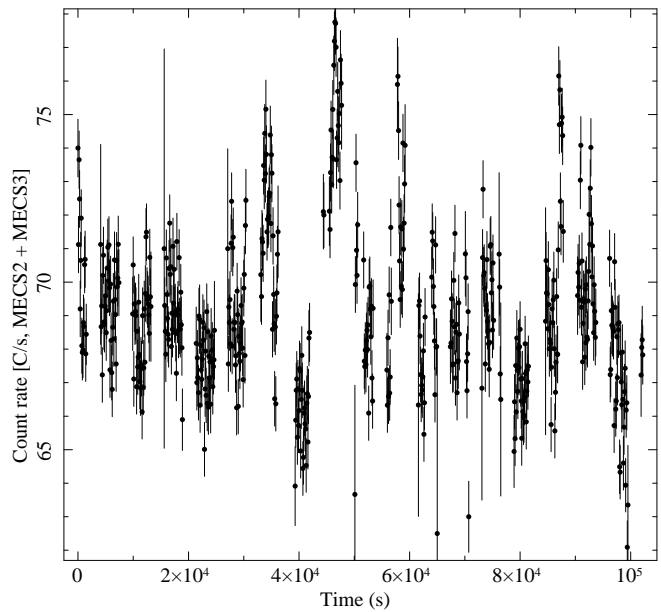}
\caption{1-10 keV MECS23  light curve. The bin time is 64 s.}
\label{figure:ME23lc}
\end{figure}
\begin{figure}
\centering
\includegraphics[scale=.65]{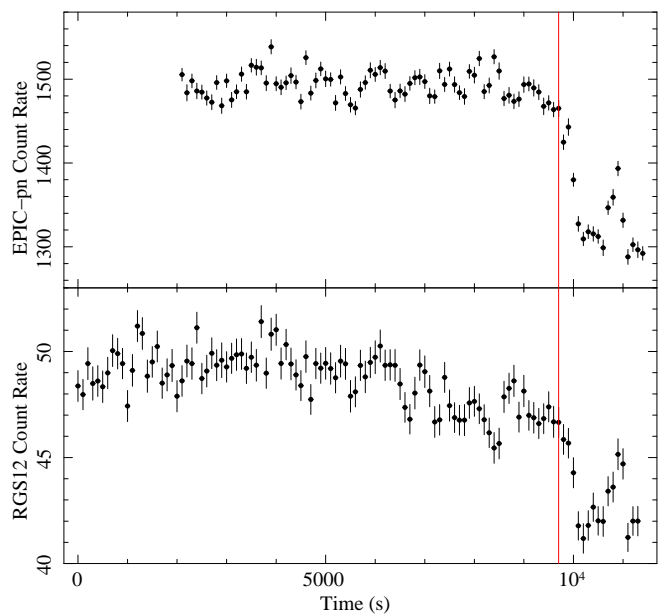}
\caption{ Top: 0.3-10 keV background-subtracted EPIC-pn light curve. Bottom: 0.35-2 keV first-order background-subtracted RGS12 light curve. The bin time is 100 s for both. The RGS and Epic-pn events above 9.7 ks from the start time (delimited by the red vertical line) were excluded from our spectral analysis.}
\label{figure:RGS12lc}
\end{figure}

The narrow field instruments (NFIs) on board the \textit{BeppoSAX} satellite observed the source between 8  April 2000 13:28:08 UTC and 9 April 2000 17:53:28 UTC (ObsId. 2083600400). The NFIs consists of four co-aligned instruments which cover three decades in energy, from 0.1 keV  to 200 keV. The Low-Energy Concentrator Spectrometer \citep[LECS, operating in the range 0.1-10 keV;][]{Parmar_97} and the Medium-Energy Concentrator Spectrometer \citep[MECS, 1.3-10 keV;][]{boella_97} have imaging capability with a field of view (FOV) of 20$\arcmin$ and 30$\arcmin$ radii, respectively. The High-Pressure Gas Scintillator Proportional Counter \citep[HPGSPC, 7-60 keV;][]{manzo_97} and the Phoswich Detector System \citep[PDS, 13-200 keV;][]{frontera_97} are non-imaging instruments, and background rejection is obtained by using the off-source data accumulated during the rocking of the collimators.
Of the three MECS modules, only MECS2 and MECS3 were active during the observation; the event files of these two instruments were merged and labelled as MECS23.
Using the {\tt XSELECT} tool, we extracted the MECS23 light curve and the LECS and MECS23 spectra from clean-event files downloaded from the Multi-Mission Interactive Archive at the ASI Space Science Data Center (SSDC). We selected the source events for LECS and MECS23 from a circular region of 6$\arcmin$ radius, centered on the source. The background events were extracted from a circular region with the same radius adopted for the source-event extractions and centered in a detector region far from the source.
The MECS23 light curves with a 64 s bin time shows a roughly constant count rate of 70 c s$^{-1}$ (see Fig. \ref{figure:ME23lc}).
We used the HP and PDS spectra made available by SSDC.
The exposure time of the LECS, MECS23, HP, and PDS spectra are 
23 ks, 50 ks, 46 ks, and 22.5 ks, respectively.
We grouped each spectrum to have a minimum of 25 counts per energy bin.
%%%%%%%%%%%%%%%%%%%%%%%

The \textit{XMM-Newton} observatory \citep{jansen_01} observed GX 9+9 on 4 September 2001 between 9:43:05 UTC and 15:17:33 UTC for a duration time of 20 ks (ObsId. 0090340101). The PN-type CCD detector of the European Photon Imaging Camera
\citep[Epic-pn;][]{struder_01} was operating in Timing mode. The Reflecting Grating Spectrometer \citep[RGS;][]{herder_01}, consisting of two modules (RGS1 and RGS2), was operating in standard spectroscopy mode (High Count Rate submode).
We reprocessed the data using the Science Analysis Software (SAS) v18.0.0, obtaining the calibrated photon event files using the {\tt epproc} and {\tt rgsproc} tools. 
The EPIC-pn count rate was, on average, higher than 1000 c s$^{-1}$; because of this, the EPIC-pn instrument often went  in burst mode, strongly reducing the net exposure time at 2.2 ks. The Epic-pn events were selected between 0.3 and 10 keV using only single- and double-pixel events (PATTERN $\le$ 4) that were optimally calibrated for spectral analysis (FLAG $=0$). 

 We extracted the EPIC-pn and RGS background-subtracted  light curves. The RGS12 light curve was  obtained combining the first-order events collected by RGS1 and RGS2 (see Fig. \ref{figure:RGS12lc}). Both  light curves show a drop in the count rate above 9.7 ks from the start time of the RGS observation. To avoid possible spectral changes during the observation, we excluded the events collected in the time interval from  9.7 ks up to the end of the observation.
 Using the \texttt{epatplot} SAS tool, we verified that Epic-pn spectrum underwent pile-up issues estimating that the pile-up fraction was higher than 13\% in the 6-10 keV energy range. We excluded the  two brightest CCD central columns to mitigate the pile-up effects with a pile-up fraction close to 10\%.
We used the columns between 20 and 36 and between 39 and 51 to extract the source spectrum, while the EPIC-pn background spectrum was obtained by the \textsc{RAWX} columns between 2 and 6, where the source contamination is negligible. 
After the pile-up correction, the response matrix and ancillary file were created following standard procedures. The spectrum was obtained adopting a spectral oversample factor of  5 and a minimum of 25 counts per energy bin.

Because of the high flux of the source, we checked whether pile-up affected the RGS spectra. We fitted the first and second order of the RGS1 and RGS2 spectra adopting an absorbed power-law component. We found that the spectra are similar to each other concluding that pile-up, if present, is moderate. 
Adopting standard procedures\footnote{\url{http://xmm2.esac.esa.int/docs/documents/CAL-TN-0075-1-0.pdf}}, we estimated that the peak flux in the 0.45-2 keV energy range is less than 0.12 photons cm$^{-2}$ s$^{-1}$ at 2 keV, corresponding to a very light pile-up.

The RGS12 spectrum was obtained by combining the first-order spectra of
RGS1 and RGS2 using the SAS tool {\tt rgscombine}.

\section{Spectral analysis}

All the spectra were grouped to have a minimum of 25 counts per energy bin. 
The adopted energy ranges are 0.3-4 keV, 1.8-10 keV, 8-25 keV, 15-40 keV, 0.35-2 keV, and 2-10 keV for LECS, MECS23, HP, PDS, RGS12, and Epic-pn, respectively. 
To fit the spectra we used XSPEC v12.10.1o; we adopted the cosmic abundances and the cross sections
derived by \cite{Wilms_00} and \cite{Verner_96}, respectively.
With the aim of accounting for the interstellar absorption, we adopted the T{\"u}bingen-Boulder model ({\sc TBabs} in XSPEC). The uncertainties are
reported at the 90\% confidence level (c.l.).
We exclude the energy range between 0.45 and 0.65 keV in the RGS12 spectrum because the high statistics does not allow us to fit the feature at 0.54 keV, which  is interpreted as a calibration issue  associated with the K edge of neutral oxygen in the RGS spectra \citep{deVries_03,Iaria_19}. For the same reason we also excluded the EPIC-pn energy range between 2.2 and 2.4 keV because the presence of the  M edge associated with neutral Au.  To take into account of the different normalizations associated with the different instruments, each model shown below  is multiplied by a constant. The value of the constant was kept fixed at 1 for the MECS23 spectrum and left free to vary for the other spectra.

Initially, we fitted the spectrum using a model composed of a blackbody component ({\sc bbody} in XSPEC) and a Comptonized component \citep[{\sc CompTT} in XSPEC,][]{Tita_94} as reported by \cite{Kong_06}. Because   the \textit{BeppoSAX} and \textit{XMM-Newton} spectra were collected at different times, we left the blackbody temperature and the interstellar absorption free to vary independently for the two spectra. The adopted model does not fit the broadband spectrum ($\chi^2$(d.o.f.)$=$ 3578(2259)); large residuals are present in the whole energy band. A similar result was obtained replacing the {\sc CompTT} component with the Comptonized component {\sc CompST} \citep{Suny_80} or {\sc nthcomp} \citep{Zdi_96} in which we assumed that the seed photons have a blackbody spectrum ({\tt inp\_type} parameter fixed to 0). 
 
 Then we adopted a model composed of a disk-blackbody component \citep[{\sc diskbb} in XSPEC;][]{Mitsuda_84} and a Comptonized component ({\sc Compbb}) \citep{Nishimura_86} as suggested by \cite{Savolainen_09} to fit the spectrum of GX 9+9. As done by the same authors, we tied the electron temperature to the seed-photon temperature of the Comptonized component.
 The model fits the data below 10 keV; however, large residuals persist at higher energies. We obtained a $\chi^2$(d.o.f.) of 3012(2260). This result suggests that the thermal component  is probably associated with  thermal emission coming from the inner region of the accretion disk and that the Comptonized component describes an optically thick plasma, contrarily to what is assumed using the {\sc Compbb} component. 

\begin{figure*}
\centering
\includegraphics[scale=.65]{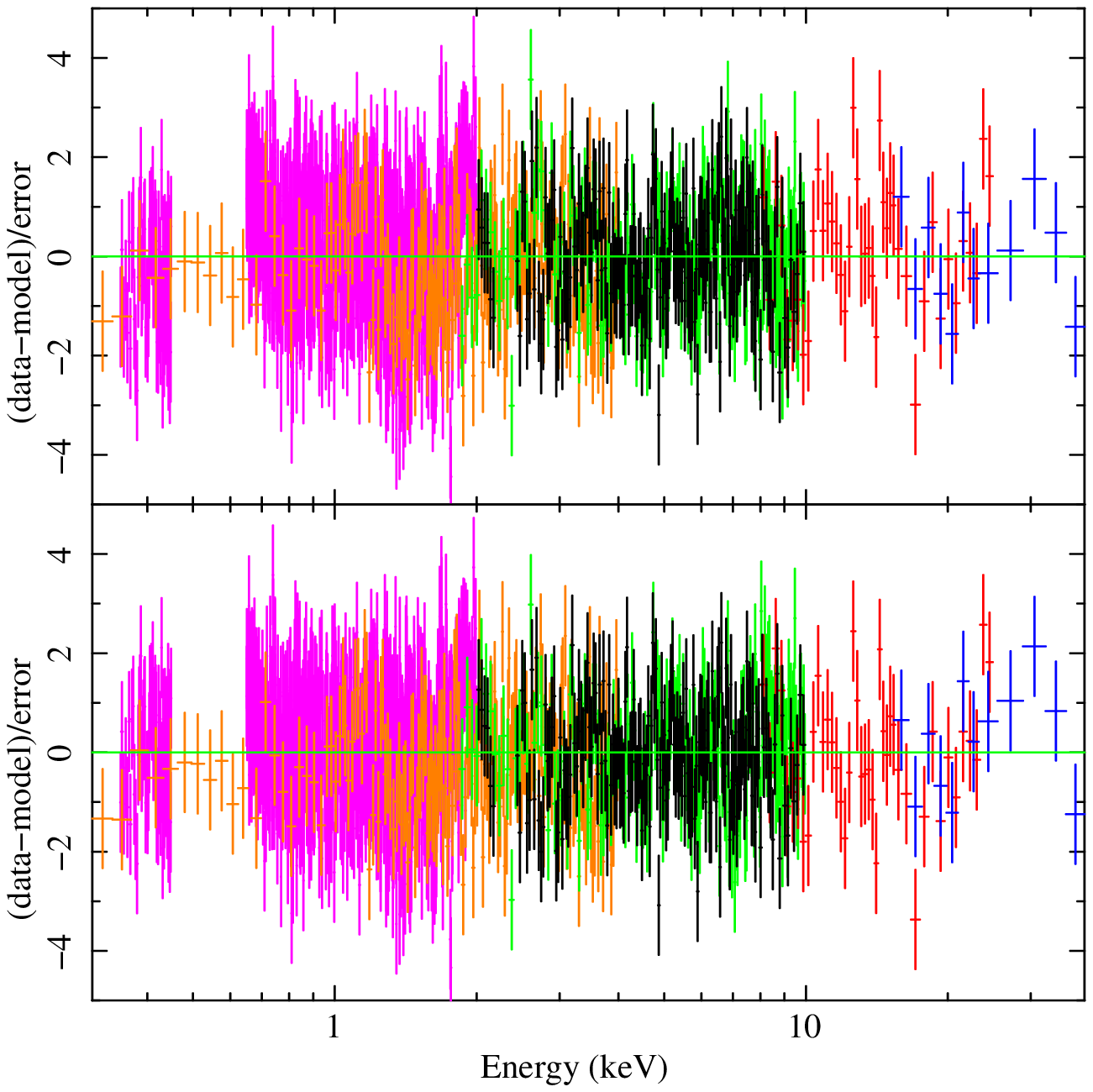}
\includegraphics[scale=.65]{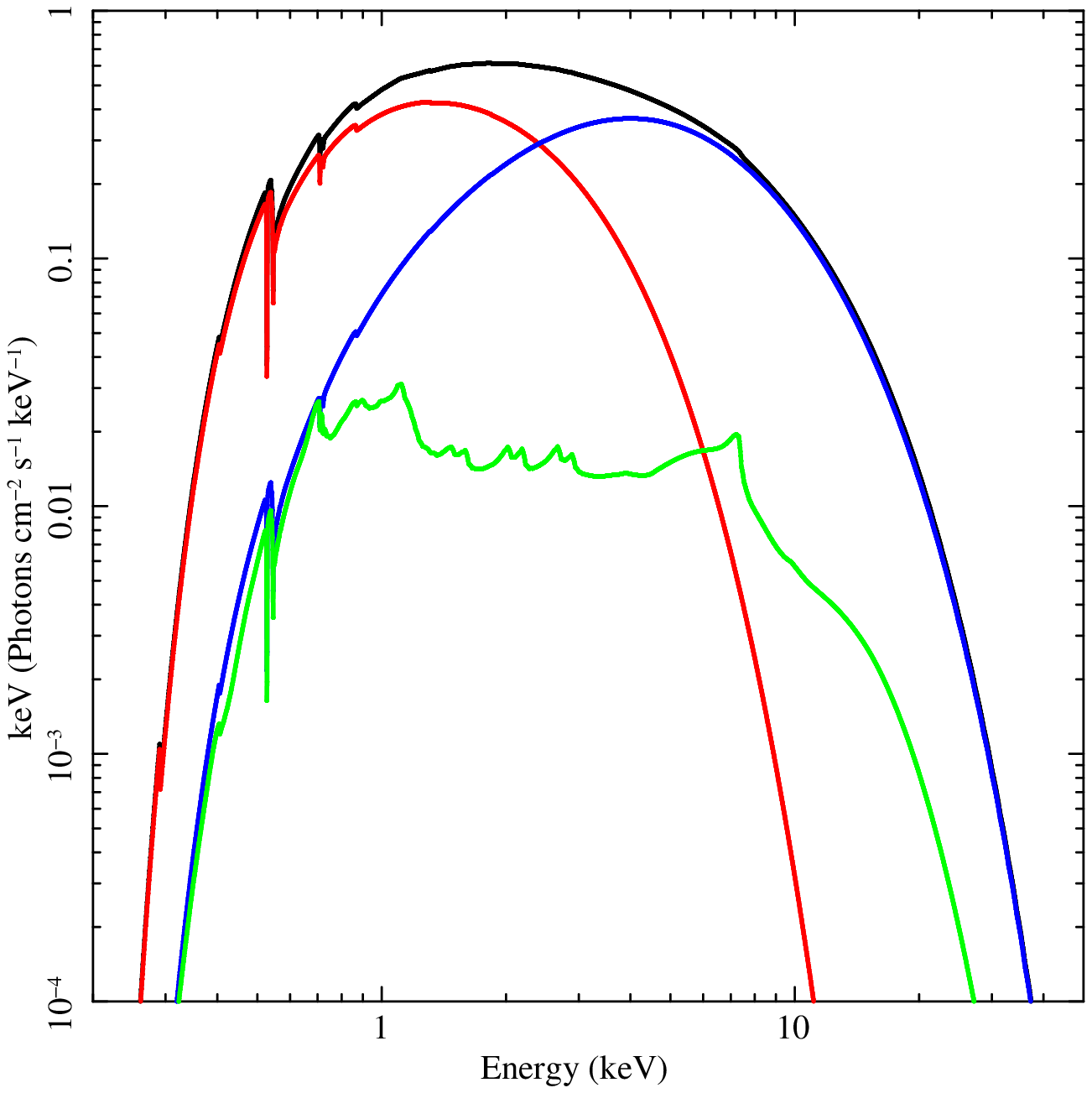}
\caption{Left panel: Residuals corresponding to {\tt Model 1} and {\tt Model 2} (top and bottom panel, respectively). The orange, magenta, black, green, red, and blue data are associated with the LECS, RGS12, Epic-pn, MECS23, HP, and PDS spectra, respectively.  Right panel: Unfolded spectrum corresponding to {\tt Model 2}. The red, blue, and green curves represent  {\sc diskbb}, {\sc nthComp,} and the reflection of the Comptonized component, respectively.}
\label{figure:res_bb_nthcomp}
\end{figure*}

In light of these previous attempts, we adopted a model composed of a disk-blackbody component {\sc diskbb} plus a Comptonized component {\sc nthcomp} in which we assume that the seed photons have a blackbody spectrum ({\tt inp\_type} parameter fixed to 0). 
 The adopted model (hereafter  {\tt Model 1}) is defined as 
$$
 \texttt{Model 1} = \textsc{TBabs*(diskbb+nthComp[0])}.
 $$
 {\tt Model 1} improves the fit above 10 keV; we found 
a $\chi^2$(d.o.f.) of 2825(2259) and an improvement of $\Delta\chi^2 \simeq 185$ with respect to the model \textsc{TBabs*(diskbb +Compbb)}.
We show the best-fit parameters in the  Col. 3 of Table \ref{tab:1}; the residuals are shown in the top left panel  of Fig. \ref{figure:res_bb_nthcomp}. 

Although the residuals seem to indicate that the model fits the spectrum overall, we observe that the RGS12 spectrum below 1 keV is not well fitted, and residuals associated with the Epic-pn spectrum are present between 3 and 5 keV. 
To further improve the fit we convolved the Comptonized component with the reflection component {\sc rxfconv} \citep[see][and references therein]{Done_06,Kole_11},  imposing  that the emission hitting the disk comes from the Comptonized component.
%The component {\sc rxfconv} has five parameters: the first is the redshift $z,$ which we kept fixed to zero, the second is the iron abundance Fe$_{\rm abundance}$ , which we kept fixed to the solar abundance, the third is the cosine of the inclination angle $\theta$, which we left free to vary, the fourth is the ionization parameter of the reflecting surface of the accretion disk log$(\xi)$, which we left free to vary, and, finally, the fifth parameter is the relative reflection normalization of the component {\sc rxfconv}, measured in units of solid angle $\Omega/2\pi$ subtended by the reflector as seen from the corona (rel$_{\rm refl}$, hereafter). 
The component {\sc rxfconv} has five
parameters:  redshift $z$ (which we kept fixed to
zero);  iron abundance Fe$_{\rm abundance}$ (kept fixed to the solar abundance);  cosine of
the inclination angle $\theta$ (left free to vary);  ionization parameter of the reflecting surface of the accretion disk log$(\xi)$ (which we left free to vary); and  the relative reflection normalization of the component {\sc rxfconv}, measured in units of the solid angle $\Omega/2\pi$ subtended
by the reflector as seen from the corona (hereafter rel$_{\rm refl}$).

Because we expect that the reflection region is close to the NS, we added the component {\sc rdblur} \citep{Fabian_89} to take into account  the smearing due to general and special relativistic
effects. {\sc rdblur} has four parameters:  inclination angle of the source;  inner and outer radius, R$_{\rm in;refl}$ and R$_{\rm out}$, of the reflection region in units of gravitational radii ($GM/c^2$, where $M$ is the NS mass); and the power-law dependence of emissivity, Betor10. Since the  value of the parameter R$_{\rm out}$ was  insensitive to the fit, we fixed the value at 3000 $GM/c^2$. In summary, the model (hereafter  {\tt Model 2})  is defined as 
\begin{equation*} 
\begin{aligned}
 {\texttt { Model 2 = }} {\textsc {TBabs*(diskbb + rdblur*rfxconv*}}\\
{ \textsc{nthComp[0])}}.
\end{aligned}
\end{equation*}
The addition of the reflection component improves the fit: we obtain a $\chi^2$(d.o.f.) of 2722(2254) and the F-test probability of chance improvement with respect to {\tt Model 1} equal to $1.4 \times 10^{-16}$. Furthermore, the addition of the reflection component improves the fitting of the RGS12 spectrum below 1 keV.
We show the best-fit values of the parameters in the  Col. 4 of Table \ref{tab:1}; the residuals and the unfolded spectrum are shown in the bottom left   and right panel of Fig. \ref{figure:res_bb_nthcomp}, respectively.
\begin{table} 
\centering
\begin{threeparttable}
\caption{Best-fit parameters for an Eastern Model scenario. The associated errors are at the 90$\%$ c.l.}
%\centering
\scriptsize
\begin{tabular}{l@{\hspace{2pt}}l@{\hspace{\tabcolsep}}c@{\hspace{\tabcolsep}}c@{\hspace{\tabcolsep}} c@{\hspace{\tabcolsep}} c@{\hspace{\tabcolsep}}}
\hline
\hline

Model & Component &\multicolumn{2}{c}{Model 1{\textit{$\rm ^a$}}}&
  \multicolumn{2}{c}{Model 2{\textit{$\rm ^b$}}} \\

  &    & SAX  & XMM & SAX & XMM  \\
\hline

{\sc Constant} &C$_{\rm RGS12}$  &  \multicolumn{2}{c}{$0.942 \pm 0.012$} & \multicolumn{2}{c}{$0.956 \pm 0.009$} \\
               & C$_{\rm LECS}$   & \multicolumn{2}{c}{$0.799 \pm 0.005$} & \multicolumn{2}{c}{$0.800 \pm 0.005$} \\
            & C$_{\rm MECS23}$   & \multicolumn{2}{c}{1 (frozen)} & \multicolumn{2}{c}{1 (frozen)} \\    
 & C$_{\rm HP}$   & \multicolumn{2}{c}{$0.999 \pm 0.005$} & \multicolumn{2}{c}{$1.007 \pm 0.005$} \\
& C$_{\rm PDS}$   & \multicolumn{2}{c}{$0.675 \pm 0.014$} & \multicolumn{2}{c}{$0.671 \pm 0.012$} \\
& C$_{\rm EPIC-pn}$   & \multicolumn{2}{c}{$0.744 \pm 0.005$} & \multicolumn{2}{c}{$0.748 \pm 0.004$} \\\\

{\sc TBabs} & N$_{\rm H}${\textit{$\rm ^c$}}  & $3.73 \pm 0.12$ & $2.18 \pm 0.02$ & 
                                    $3.61 \pm 0.09 $ & $2.20 \pm 0.02$\\\\             

{\sc diskbb}  & k$T$ (keV)&  $0.82\pm0.03$ &$0.78 \pm 0.03$&
                              $ 0.86^{+0.08}_{-0.02}$&$0.82\pm0.02 $\\
  
 & R${_{\rm in}}\sqrt{\cos{\theta}}$ (km){\textit{$ \rm{^d}$}}  & \multicolumn{2}{c}{$9.0 \pm 0.6$} & 
           \multicolumn{2}{c}{$8.1^{+0.2}_{-1.1}$} \\\\

{\sc nthComp} & $\Gamma$ & \multicolumn{2}{c}{$2.37\pm0.11 $ } & 
                           \multicolumn{2}{c}{$2.22\pm 0.05 $} \\ 
  
 & kT$_{\rm e}$ (keV) & \multicolumn{2}{c}{$2.95 \pm 0.10$}  &
                        \multicolumn{2}{c}{$2.80^{+0.09}_{-0.04}$}\\ 
                        
 & kT$_{\rm bb}$ (keV) & \multicolumn{2}{c}{$1.18 \pm 0.06$}  &
                         \multicolumn{2}{c}{$1.14^{+0.10}_{-0.07}$} \\ 
                         
&inp$\_$type& \multicolumn{2}{c}{0}& 
              \multicolumn{2}{c}{0}\\

 & Norm ($\times 10^{-2}$) & \multicolumn{2}{c}{$7.4 \pm 0.8$}&
          \multicolumn{2}{c}{$5.53 ^{+0.04}_{-0.07}$}\\\\

{\sc rdblur} & Betor10 & \multicolumn{2}{c}{-}&
                         \multicolumn{2}{c}{$<-3.7$} \\ 
                         
 & R$_{\rm in;refl}$ $\rm (GM/c^2)$ & \multicolumn{2}{c}{-}& 
                                 \multicolumn{2}{c}{$10^{+2}_{-4}$}  \\

 & R$_{\rm out}$ $\rm (GM/c^2)$ & \multicolumn{2}{c}{-}& 
                                  \multicolumn{2}{c}{$3000$ (frozen)}  \\

 & $\theta$ (deg) & \multicolumn{2}{c}{-}& 
                    \multicolumn{2}{c}{$43^{+6}_{-4}$} \\\\ 

{\sc rfxconv} & rel$_{\rm refl}$& \multicolumn{2}{c}{-}& 
                                  \multicolumn{2}{c}{$0.18 \pm 0.04 $} \\
                                  
 & Fe$_{\rm abund}$ & \multicolumn{2}{c}{-}& 
                      \multicolumn{2}{c}{1 (frozen)}\\
                     
 & $\cos \theta$$\rm \; ^e$ & \multicolumn{2}{c}{-}& \multicolumn{2}{c}{-}\\
 
 & log($\xi$) & \multicolumn{2}{c}{-}& 
                \multicolumn{2}{c}{$2.694^{+0.012}_{-0.089} $}  \\\\

 & $\chi^2/dof$ & \multicolumn{2}{c}{2825/2259} & 
                   \multicolumn{2}{c}{2722/2254}\\ 
\hline
\hline
\end{tabular}
      \begin{tablenotes}
 \item[a] Model 1 = {\sc TBabs*(diskbb+nthComp[0])},
\item[b] Model 2 = {\sc TBabs*(diskbb+rdblur*rfxconv*nthComp[0])}, 
\item[c] \textrm{equivalent  column density of neutral hydrogen in units of 10$^{21}$ atoms cm$^{-2}$},
\item[d] {\textrm{assuming a distance to the source of 5 kpc}},
\item[e] \textrm{this parameter is tied to the parameter $\theta$ of the {\sc rdblur} component}.
 \end{tablenotes}
\label{tab:1}
 \end{threeparttable}
\end{table}

  We found that the equivalent column density of neutral hydrogen, $N_{\rm H}$, has a higher value of $(3.61 \pm 0.09) \times 10^{21}$ during the \textit{BeppoSAX} observation, while it is $ (2.20 \pm 0.02) \times 10^{21}$ cm$^{-2}$ during the \textit{XMM-Newton} observation. The disk-blackbody temperature k$T$ is slightly higher during the \textit{BeppoSAX} observation; k$T$ is $\sim 0.86$ keV and $\sim 0.82$ keV during the \textit{BeppoSAX} and \textit{XMM-Newton} observation, respectively. The seed-photon temperature k$T_{\rm bb}$ and the electron temperature k$T_{e}$ of the corona are $1.14^{+0.10}_{-0.07}$ keV and $2.80^{+0.09}_{-0.04}$ keV, respectively. Finally, the power-law photon index, $\Gamma $, is $2.22 \pm 0.05$.
  
  The ionization parameter log($\xi$) inferred from the reflection component is $\sim 2.70$ and the reflection component has a relative normalization of $0.18 \pm 0.04$.
  The inner  radius of the reflecting region is $10^{+2}_{-4}$ gravitational radii; the power-law dependence of emissivity, Betor10, is less than $-3.7$. Finally, we constrained the inclination angle of the system, finding $43^{+6}_{-4}$ degrees. 
    \begin{table} [!htbp]
\centering
\begin{threeparttable}
\caption{Best-fit parameters for a Western  Model scenario. The associated errors are at the 90$\%$ c.l.}
%\centering
\scriptsize 
\begin{tabular}{l@{\hspace{2pt}}l@{\hspace{\tabcolsep}}c@{\hspace{\tabcolsep}}c@{\hspace{\tabcolsep}} c@{\hspace{\tabcolsep}} c@{\hspace{\tabcolsep}}}
\hline
\hline

Model & Component &\multicolumn{2}{c}{Model 3{\textit{$\rm ^a$}}}&
  \multicolumn{2}{c}{Model 4{\textit{$\rm ^b$}}}\\

  &    & SAX  & XMM & SAX & XMM  \\
\hline

{\sc Constant} &C$_{\rm RGS12}$  &  \multicolumn{2}{c}{$0.743 \pm 0.005$} & \multicolumn{2}{c}{$0.805 \pm 0.011$} \\
               & C$_{\rm LECS}$   & \multicolumn{2}{c}{$0.944 \pm 0.011$} & \multicolumn{2}{c}{$0.801 \pm 0.005$} \\
            & C$_{\rm MECS23}$   & \multicolumn{2}{c}{1 (frozen)} & \multicolumn{2}{c}{1 (frozen)} \\    
 & C$_{\rm HP}$   & \multicolumn{2}{c}{$0.999 \pm 0.006$} & \multicolumn{2}{c}{$1.005 \pm 0.006$} \\
& C$_{\rm PDS}$   & \multicolumn{2}{c}{$0.672 \pm 0.013$} & \multicolumn{2}{c}{$0.676 \pm 0.014$} \\
& C$_{\rm EPIC-pn}$   & \multicolumn{2}{c}{$0.655 \pm 0.014$} & \multicolumn{2}{c}{$0.618 \pm 0.012$} \\\\

{\sc TBabs} & N$_{\rm H}${\textit{$\rm ^c$}}  & $3.80 \pm 0.15$ & $2.20 \pm 0.03$ &      $3.92 \pm 0.13 $ & $2.26 \pm 0.03$\\      \\             

{\sc bbodyrad}  & k$T$ (keV)&  $1.49\pm0.07$ &$1.65\pm 0.06$&
                              
                              $ 1.79^{+0.09}_{-0.18}$& $1.85^{+0.07}_{-0.15}$\\
  
 & R${_{\rm in}}$ (km){\textit{$ \rm{^d}$}}  & \multicolumn{2}{c}{$2.13 \pm 0.13$} & 
           
           \multicolumn{2}{c}{$1.50 \pm 0.15$}\\\\

{\sc nthComp} & $\Gamma$ & \multicolumn{2}{c}{$1.94\pm0.04 $ } &  \multicolumn{2}{c}{$2.10\pm 0.15 $} \\ 
  
 & kT$_{\rm e}$ (keV) & \multicolumn{2}{c}{$2.77 \pm 0.08$} &
                        \multicolumn{2}{c}{$2.9 \pm 0.2$} \\ 
                        
 & kT$_{\rm bb}$ (keV) & $0.69 \pm 0.03$  &$0.71 \pm 0.03$ &
                         $0.87\pm 0.07$ & $ 1.01 \pm 0.08$ \\

&inp$\_$type& \multicolumn{2}{c}{1}& 
              \multicolumn{2}{c}{1}\\

 & Norm  & \multicolumn{2}{c}{$0.830 \pm 0.012$}&
          \multicolumn{2}{c}{$0.78 \pm 0.02 $}\\\\
          
{\sc rdblur} & Betor10 & \multicolumn{2}{c}{-}&
                         
                         \multicolumn{2}{c}{$-2.4^{+0.2}_{-0.3}$}\\ 
                         
 & R$_{\rm in;refl}$ $\rm (GM/c^2)$ & \multicolumn{2}{c}{-}& 
                                 \multicolumn{2}{c}{$<15$} \\

 & R$_{\rm out}$ $\rm (GM/c^2)$ & \multicolumn{2}{c}{-}& 
                                  \multicolumn{2}{c}{$>790$}\\

 & $\theta$ (deg) & \multicolumn{2}{c}{-}& 
                    \multicolumn{2}{c}{$51^{+9}_{-2}$} \\\\ 

{\sc rfxconv} & rel$_{\rm refl}$& \multicolumn{2}{c}{-}& 
                    \multicolumn{2}{c}{$0.21 \pm 0.03 $} \\
                                  
 & Fe$_{\rm abund}$ & \multicolumn{2}{c}{-}& 
                      \multicolumn{2}{c}{1 (frozen)}\\
                     
 & $\cos \theta$$\rm \; ^e$ & \multicolumn{2}{c}{-}& \multicolumn{2}{c}{-}  \\
 
 & log($\xi$) & \multicolumn{2}{c}{-}& 
                \multicolumn{2}{c}{$2.39^{+0.07}_{-0.04}$}\\\\

 & $\chi^2/dof$ & \multicolumn{2}{c}{2842/2258} & 
                   \multicolumn{2}{c}{2680/2252}\\ 
\hline
\hline
\end{tabular}
      \begin{tablenotes}
 \item[a] Model 3 = {\sc TBabs*(bbodyrad+nthComp[1])},
\item[b] Model 4 = {\sc TBabs*(bbodyrad+rdblur*rfxconv*nthComp[1])}, 
\item[c] \textrm{equivalent column density of neutral hydrogen in units of 10$^{21}$ atoms cm$^{-2}$},
\item[d] {\textrm{assuming a distance to the source of 5 kpc}},
\item[e] \textrm{this parameter is tied to the parameter $\theta$ of the {\sc rdblur} component}.
 \end{tablenotes}
\label{tab:2}
 \end{threeparttable}
\end{table}

  We obtain an absorbed flux in the 0.3-40 keV energy range of $6.8 \times 10^{-9}$ erg s$^{-1}$ cm$^{-2}$. We extrapolated an 0.1-100 keV unabsorbed total flux of $7.6 \times 10^{-9}$ erg s$^{-1}$ cm$^{-2}$; the Comptonized and disk-blackbody components have an unabsorbed flux of $5.0 \times 10^{-9}$ and  $2.4 \times 10^{-9}$ erg s$^{-1}$ cm$^{-2}$, respectively. Assuming a distance to the source of 5 kpc, the 0.1-100 luminosity of GX 9+9 is $2.3 \times 10^{37}$ erg s$^{-1}$, i.e., around 10\% of the Eddington luminosity for a NS mass of 1.4 M$_\odot$.
\begin{figure*}
\centering
\includegraphics[scale=.65]{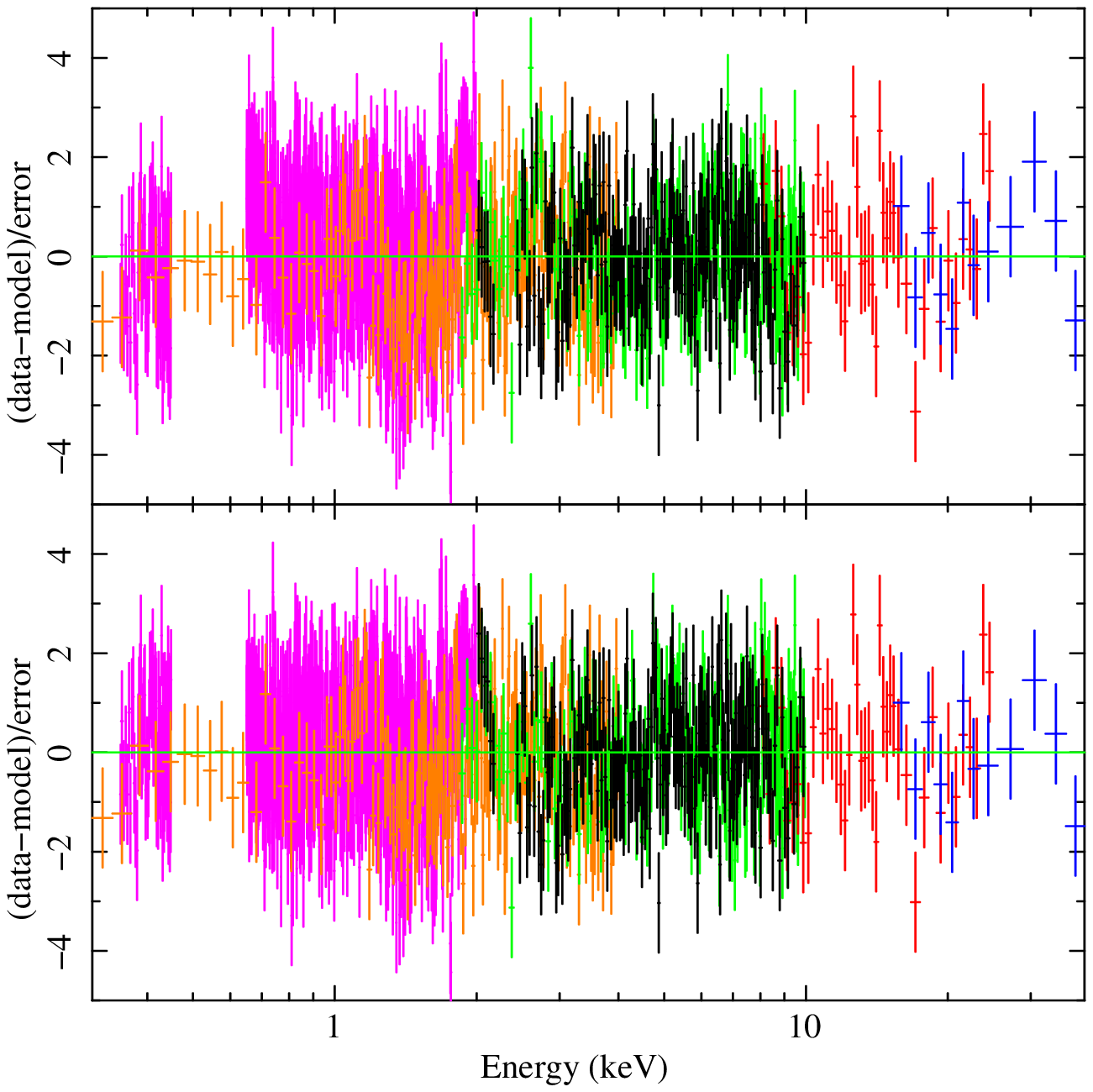}
\includegraphics[scale=.65]{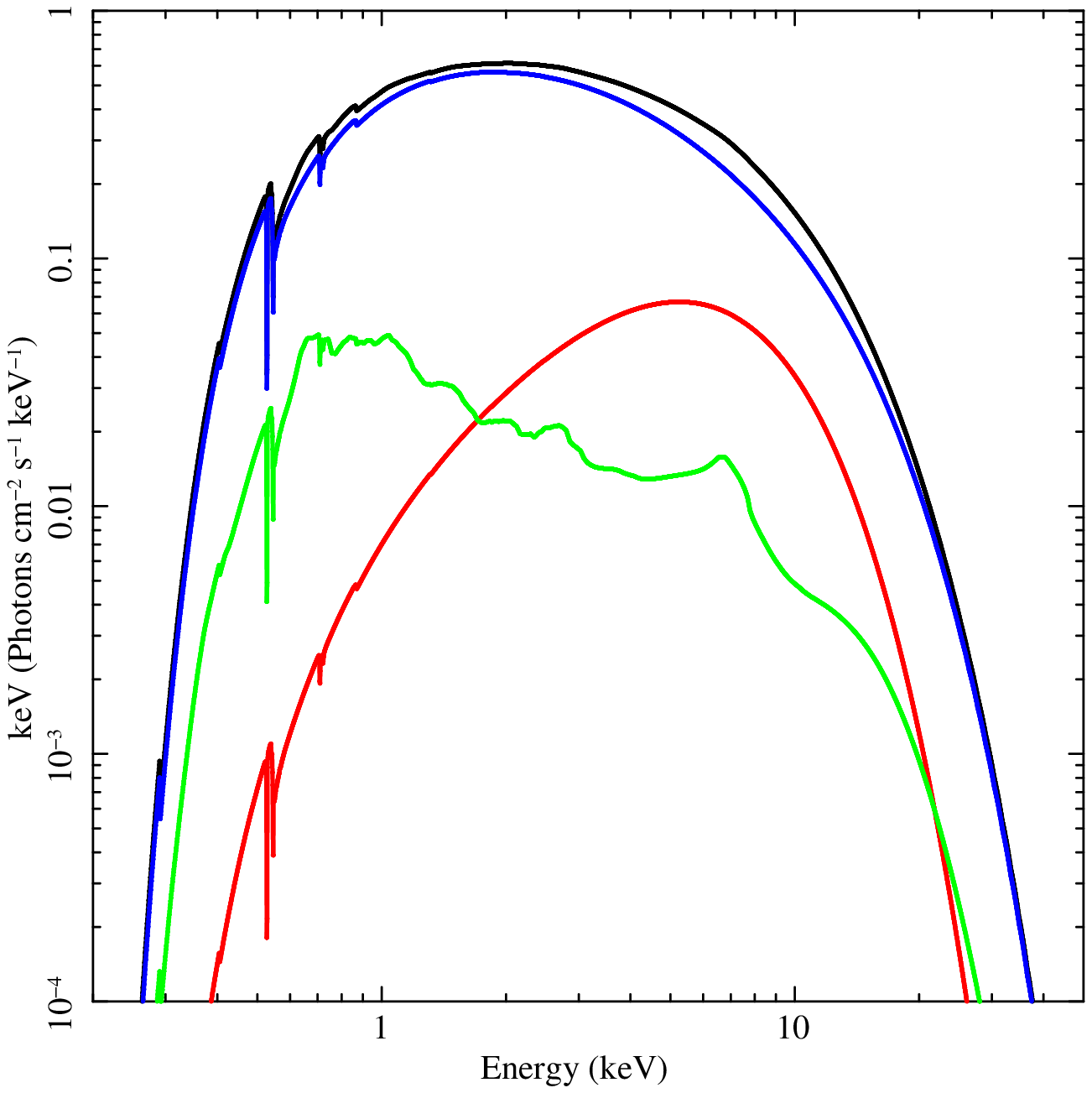}
\caption{Left panels:  Residuals corresponding to {\tt Model 3} and {\tt Model 4} (top and bottom panel, respectively). The colors are defined  in Fig. \ref{figure:res_bb_nthcomp}.  Right panel: Unfolded spectrum corresponding to {\tt Model 4}. The blue, red, and green  curves represent  {\sc nthComp}, {\sc bbodyrad,} and the smeared reflection of the Comptonized component, respectively.}
\label{figure:res_bb_nthcomp2}
\end{figure*}

{\tt Model 2} describes an Eastern Model scenario; moving forward, we also investigated  the possibility that the broadband spectrum could be described by a Western Model scenario in which the thermal components mimics a saturated Comptonization. We imposed that the Comptonized component {\sc nthcomp} originates from seed photons with a disk-blackbody spectrum ({\tt inp$\_$type}=1) and added a blackbody component ({\sc bbodyrad} in XSPEC) to mimic a saturated Comptonization. We left free to vary independently the seed-photon temperature and the blackbody temperature associated with the \textit{BeppoSAX} and \textit{XMM-Newton} spectra. The model (hereafter   {\tt Model 3}) is defined as
$$
 \texttt{Model 3} = \textsc{TBabs*(bbodyrad+nthComp[1])}.
$$
 By fitting the spectrum we found 
a $\chi^2$(d.o.f.) of 2842(2258).  
We show the best-fit parameters in the  Col. 3 of Table \ref{tab:2}; the residuals are shown in the top left  panel of  Fig. \ref{figure:res_bb_nthcomp2}. Residuals are evident in the RGS12 spectrum below 1 keV. As previously done,  
we added a smeared reflection component assuming that the only incident emission is that associated with the Comptonized component. The model (hereafter  {\tt Model 4}) is defined as 
\begin{equation*} 
\begin{aligned}
 {\texttt { Model 4 = }} {\textsc {TBabs*(bbodyrad + rdblur*rfxconv*}}\\
{ \textsc{nthComp[1])}}.
\end{aligned}
\end{equation*}
The addition of the reflection component improved the fit: we obtained a $\chi^2$(d.o.f.) of 2680(2252) and the F-test probability of chance improvement with respect to {\tt Model 3} equal to $4 \times 10^{-26}$. 
We show the best-fit values of the parameters in the  Col. 4 of Table \ref{tab:2}. The residuals and the unfolded spectrum are shown in the bottom left  and right panel of  Fig. \ref{figure:res_bb_nthcomp2}, respectively.  

 We found that the equivalent column density of neutral hydrogen, $N_{\rm H}$, has a higher value of $(3.92 \pm 0.13) \times 10^{21}$ during the \textit{BeppoSAX} observation, while it is $ (2.26 \pm 0.03) \times 10^{21}$ cm$^{-2}$ during the \textit{XMM-Newton} observation. The blackbody temperature (k$T$) and the seed-photons temperature (k$T_{\rm bb}$) are slightly higher  during the \textit{XMM-Newton} observation although they are compatible each other within the 90\% c.l. During the \textit{BeppoSAX}
and \textit{XMM-Newton} observation  k$T$ is $\sim 1.79$ keV and $\sim 1.85$ keV, and k$T_{\rm bb}$ is $\sim 0.87$ keV and $\sim 1.01$ keV, respectively. 
 The electron temperature k$T_{e}$ of the corona and the photon index $\Gamma$ are $2.9 \pm 0.2$ keV and $2.10 \pm 0.15$, respectively.
 
 The ionization parameter log($\xi$) inferred from the reflection component is $\sim 2.39$ and the relative normalization of the same component is $0.21 \pm 0.03$. The inner and outer radius of the reflecting region are less than 15 and larger than 790 gravitational radii, respectively; the power-law dependence of emissivity is $-2.4^{+0.2}_{-0.3} $. Finally, we were able to constrain the inclination angle of the system, finding a value of $51^{+9}_{-2}$ degrees. 

 We obtained an absorbed flux in the 0.3-40 keV energy range of $6.9 \times 10^{-9}$ erg s$^{-1}$ cm$^{-2}$, while the extrapolated 0.1-100 keV unabsorbed flux is $7.8 \times 10^{-9}$ erg s$^{-1}$ cm$^{-2}$. The Comptonized, the blackbody, and the reflection components have an unabsorbed flux of $6.6 \times 10^{-9}$, $0.9 \times 10^{-9}$, and $0.4 \times 10^{-9}$ erg s$^{-1}$ cm$^{-2}$, respectively. Assuming a distance to the source of 5 kpc, the 0.1-100 keV luminosity of GX 9+9 is $2.3 \times 10^{37}$ erg s$^{-1}$, or around 10\% of the Eddington luminosity for a NS mass of 1.4 M$_\odot$.

It is important to mention that we also tried  the relativistic self-consistent reflection model
{\sc relxillCp} v.1.3.5 \citep{Garcia_14}. Unfortunately, this component applied to our spectrum 
gives an unrealistic reflected spectrum stronger than the incident one at energies higher than 20 keV. 
A possible explanation is that the seed-photon temperature of the Comptonization component is higher than 1 keV, while {\sc relxillCp} works with a value of k$T_{\rm bb}$ kept fixed at 0.05 keV. We therefore decided not to discuss this model further.
%(Garc{\'\i}a, private communication). 
 
 Finally, we explored the possibility that the soft excess observed in RGS12 spectra can be fitted with  a bremsstrahlung component instead of a reflection component. We added the component {\sc Bremss} to {\tt Model 1}. We obtained a $\chi^2$(d.o.f.) of 2756(2257), and the F-test probability of chance improvement with respect to {\tt Model 1} is $7 \times 10^{-13}$.  We found that the bremsstrahlung temperature is only $0.31 \pm 0.03$ keV. The total  absorbed flux in the 0.3-40 keV energy range of $6.8 \times 10^{-9}$ erg s$^{-1}$ cm$^{-2}$, while the extrapolated 0.1-100 keV unabsorbed flux is $8.7 \times 10^{-9}$ erg s$^{-1}$ cm$^{-2}$. The Comptonized, the disk-blackbody, and the bremsstrahlung  components have an unabsorbed flux of $5.3 \times 10^{-9}$, $2.7 \times 10^{-9}$, and $1.1 \times 10^{-9}$ erg s$^{-1}$ cm$^{-2}$, respectively. Assuming a distance to the source of 5 kpc, the 0.1-100 keV luminosity of GX 9+9 is $2.6 \times 10^{37}$ erg s$^{-1}$, or around 10\% of the Eddington luminosity for a NS mass of 1.4 M$_\odot$.

\section{Discussion}
We show the first broadband spectral analysis of the bright Atoll source GX 9+9 in the 0.3-40 keV energy band using \textit{BeppoSAX} and \textit{XMM-Newton} data. The broadband analysis allows us to exclude that the spectrum of GX 9+9 could be fitted by a disk-blackbody component  plus a Comptonized blackbody component {(\sc CompBB}) representing a spreading layer close the NS surface, as was suggested, for instance, by \cite{Savolainen_09} using RXTE/PCA and INTEGRAL spectra in the 4-40 keV energy range.

We found that direct emission can be fitted by two different models: the first  is composed of a disk-blackbody component plus a Comptonization having a blackbody seed-photon spectrum \citep[{\tt Model 2}, called the Eastern Model; see][]{Mitsuda_84}; the second one ({\tt Model 4}) is composed of a blackbody component plus a Comptonization having a disk-blackbody seed-photon spectrum \citep[called the   Western Model; see][]{White_88}.

In both of the models the unabsorbed bolometric flux of the reflection component is $4 \times 10^{-10}$ erg s$^{-1}$ cm$^{-2}$; the inner and outer radii of the reflecting region are not well constrained in both the cases. We found that the inner reflection radius, $R_{\rm in;refl}$, is $10^{+2}_{-4}$ and 
$<15$ gravitational radii for {\tt Model 2} and {\tt Model 4}, respectively.  The outer reflection radius $R_{\rm out}$ was fixed at  $3000$ gravitational radii for {\tt Model 2}, while it is   $>790 $ gravitational radii for {\tt Model 4}. The values found for the inclination angle $\theta$  are well constrained in the fits with both models (i.e., $43^{+6}_{-4}$ and $51 ^{+9}_{-2}$  degrees, respectively) and are also compatible with the upper limit of 63$^{\circ}$ inferred by \cite{Hertz_88}. Finally, the ionization parameter of the reflection skin above the disk is $2.694^{+0.012}_{-0.089}$ and $2.39 ^{+0.07}_{-0.04} $ for {\tt Model 2} and {\tt Model 4}, respectively, which are marginally compatible with each other within 3$\sigma$. 

Initially, we discuss the scenario in which the direct emission comes from the disk (soft thermal component) and from a Comptonized region (hard component) surrounding the NS ({\tt Model 2}). The seed photons pumped in the Comptonized region have a blackbody spectrum, suggesting that they probably  originate from the NS surface or the boundary layer. 
The thermal disk-blackbody has a temperature close to 0.85 keV, and the Comptonized component has an electron temperature of 2.8 keV, a photon index of 2.2, and a seed-photon temperature close to 1.1 keV. 
Using  Eq. 2 reported by \cite{Zdi_96}, we estimated the optical depth of the Comptonized region to be $\tau= 8.9\pm0.4$. 
The estimated Comptonization parameter  (i.e., $y=4kT_e\tau^2/(m_ec^2)$) is $y= 1.7\pm0.2$. Therefore, we estimated the seed-photon radius R$_{\rm seed}$ using the relation ${\rm R_{seed}} = 3 \times 10^4\;d [{\rm F_{Compt}}/(1 + y)]^{1/2}(kT_{\rm bb})^{-2}$ \citep{Zand_99}, where $d$ is the distance to the source in units of kpc, k$T_{\rm bb}$ is the seed-photon temperature in units of keV, and F$_{\rm Compt}$ is the bolometric flux of the Comptonized component. We found that ${\rm R_{seed}} =5.0\pm0.9$ km which implies that the Comptonized region is a limited region of the NS surface, such as an optically thick boundary layer or a portion of the NS illuminated by the inner disk. 

Since the inclination angle of the system obtained by the reflection component is $43^{+6}_{-4}$ degrees,  we estimated the apparent inner radius of the accretion disk:  $R_{\rm in} = 9.5 \pm 1.3$ km. Given the observed bolometric luminosity is 10\% of the Eddington luminosity, we corrected R$_{\rm in}$ to find the real inner radius of the disk, using the relation $r_{\rm disk} \simeq f^2 R_{\rm in} $ with $f \simeq 1.7$ \citep{Shimura_95}; we obtained $r_{\rm disk} = 27 \pm 4$ km.
We note that this value is compatible with the value found for the inner radius of the reflecting skin above the accretion disk, i.e., $21^{+4}_{-8}$ km under the assumption of a 1.4 M$_\odot$ NS mass.

Assuming that the boundary layer goes from the NS surface up to the inner radius of the disk, its geometrical thickness $l$ is roughly 11 km. Since the optical depth of the Comptonized corona is $\tau \simeq 8.9$, we estimated the order of magnitude of the electron density in the boundary layer in the hypothesis that this is identified with the corona and using the relation $\tau =n_e \sigma_T l$, where $\sigma_T$ is the Thomson cross section. We obtained $n_e \simeq 1.2 \times 10^{19}$ cm$^{-3}$. 
We also estimated the electron density $\sc{N}_e$ of
the reflecting skin above the disk using the relation $\xi = L_x/(\sc{N}_er^2)$, where $L_x$ is the unabsorbed incident luminosity between 0.1 and 100 keV, $\xi$ is the ionization parameter, and $r$ is the inner radius of the disk where the reflection component originates.
In our case $L_x \simeq 1.5 \times  10^{37}$ erg s$^{-1}$ is the luminosity of the Comptonized emission. Using the value of log($\xi$)$\simeq 2.7$ obtained from the best-fit and adopting $r=21$ km, the best-fit value of the inner radius of the reflecting region, we obtained that $\sc{N}_e \simeq 7 \times 10^{21}$ cm$^{-3}$, which is at least a factor of one hundred larger than the value obtained for the boundary layer. 
We found a reflection amplitude rel$_{\rm refl}$, which is a measure of the solid angle subtended by the reflector as seen from the Comptonizing corona, of $0.18 \pm 0.04$. 
Typical values of the reflection amplitude $\Omega/2 \pi$ for NS-LMXB atoll sources are within 0.2-0.3 \citep[e.g.,][]{disalvo_19,Marino_19,Matranga_17b,Disalvo_15,Egron_13,dai_10}. These values describe a scenario in which a spherical corona is present in the inner part of the accretion disk \citep[see Fig. 5 in][]{Dove_97}.

Finally we note that the bolometric flux of the disk corrected by the inclination angle of 43$^\circ$ is $3.3 \times 10^{-9}$ erg cm$^{-2}$ s$^{-1}$, which is similar to the value of flux obtained from the Comptonized component. This result is in agreement with the prediction of \cite{sunayev_86}, who suggested that the luminosity from the boundary layer and from the disk should be the same for a slowly spinning NS having weak magnetic field. 
\begin{figure}[ht!]
\centering
\includegraphics[scale=.65]{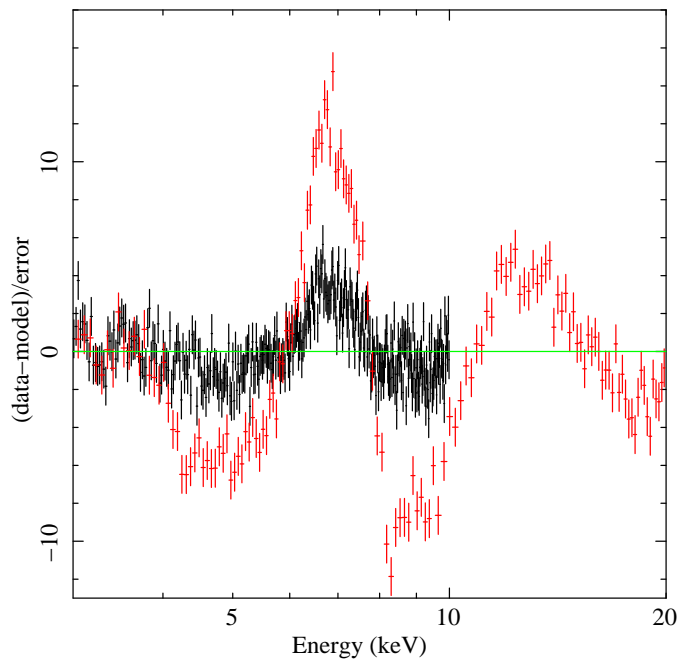}
\caption{Residuals corresponding to a simulated eXTP/SFA (black) and eXTP/LAD (red) spectrum of {\tt Model 4} after removing the smeared reflection component. The presence of a broad emission line is evident in the Fe-K region. The exposure time of the simulated spectra is 20 ks.}
\label{figure:del_line}
\end{figure} 
 
%{\tt Model 6} paints a scenario in which the Comptonized region is furnished of seed-photons having a disk-blackbody spectrum, 
On the other hand, {\tt Model 4} paints a scenario in which the seed photons entering the Comptonized region have a  disk-blackbody spectrum,
while the thermal component mimics a saturated Comptonization describing an optically thick boundary layer.
This scenario is apparently similar to what is proposed for GX 9+9 by \cite{Kong_06}, who analyzed RXTE and PCA spectra in the 2.5-20 keV energy range. However, \citeauthor{Kong_06} suggested that the Comptonized component describes the emission of an extended corona, while {\tt Model 4} includes a statistically significant reflection component with $\Omega/2 \pi=0.21 \pm 0.03$. As discussed above, this value of solid angle is compatible with a compact corona close the NS \citep{Dove_97}. In this case, a possible geometry of the accretion could be similar to what is discussed by \cite{Mainardi_10}. The authors fitted the INTEGRAL spectra of GX 5–1, GX 349+2, GX 13+1, GX 3+1, and GX 9+1 adopting two Comptonized components. The first component, describing the dominant part of the spectrum, is interpreted as thermal Comptonization of soft seed photons (${\rm kT_{bb}}<1$ keV) from the accretion disk and electron temperature between 3 and 5 keV, while the second component could describe a bulk plus thermal Comptonization of blackbody seed photons or a simple blackbody emission. The seed photons of the latter component, with temperature higher than 1 keV and corresponding radius less than 12 km, likely coming  from the NS and the innermost part of the system, are Comptonized by matter in a converging flow, dubbed the transition layer. In particular, \cite{Mainardi_10} found that the second Comptonized component is well described by a blackbody component for GX 9+1, GX 349+2, and GX 5-1 (for the last source, and  only when the hardness ratio is high). 

We found that the Comptonized component has an electron temperature of 2.9 keV, photon index of 2.1 and seed-photon temperature close to 0.9-1 keV. 
Using  Eq. 2 reported by \cite{Zdi_96} we estimated that the optical depth of the Comptonized region is $\tau= 9.4^{+1.5}_{-1.1}$. 
The estimated Comptonization parameter is $y=2.0\pm0.8$. Therefore, we estimated the seed-photon radius using the relation shown above \citep{Zand_99}. We found   ${\rm R_{seed}} = 6.9\pm1.5$ km assuming that the seed photons come from a spherical surface with radius ${\rm R_{seed}}$. On the other hand, if we assume that seed photons come from an annulus of the disk at a radius ${\rm R}$ and thickness ${\rm \Delta R}$, we can rewrite the expression as ${\rm R \Delta R = R^2_{seed}}$, not taking into account a possible correction due to the inclination angle of the source. Arbitrarily assuming that ${\rm R}$ is 15 km, we obtain a plausible value of ${\rm \Delta R = 3.2\pm 1.4}$ km. 

The blackbody radius associated with the saturated Comptonization is only 1.5 km and the unabsorbed flux associated with this component is only 12\% of the total value. A similar ratio is observed by \cite{Mainardi_10} for GX 9+1, while \cite{Kong_06} find a ratio of 30\% for GX 9+9. Since we expect that the contribution of the two components is similar \citep{sunayev_86}, we have to invoke some mechanism to explain the partial suppression of the boundary layer emission. Two explanations are reported in the literature. The first, proposed by \cite{White_88}, suggests that if
the NS spin period is close to equilibrium with the Keplerian period of the inner edge of the accretion disk, then a bit of energy is given up in the boundary layer. The second explanation, proposed by \cite{vanderklis_87}, suggests that part of the boundary layer emission is blocked by the inner disk corona or thickened disk. We note that the second mechanism is compatible with the scenario proposed by \cite{Mainardi_10} and \cite{Seifina_12}. 

For {\tt Model 4} the electron density of the reflecting skin above the disk is $\sc{N}_e >8.5 \times 10^{21}$ cm$^{-3}$ using the unabsorbed incident luminosity of the Comptonized component ($L_x \simeq 2 \times 10^{37}$ erg s$^{-1}$), log($\xi$)$=2.39 $, and  $r=31$ km, i.e.,  the upper value of the inner radius of the reflecting region.

In conclusion, our analysis shows that although we do not have   clear evidence of a broad emission line in the Fe-K region, a relativistic smeared reflection component is necessary to fit the 0.3-40 keV broadband spectrum. The lack of residuals associated with a broad emission line is likely due to the available low statistics in the Fe-K region, unlike the source GX 5-1 for which the lack of the broad emission line is explained as being due to the high ionization parameter  \citep[see][]{Homan_18}. 
Our prediction will be confirmed in the near future, after the launch of the enhanced X-Ray Timing and Polarimetry 
\citep[eXTP;][]{Derosa_19}
observatory. The Spectroscopic Focusing Array (SFA; 0.5-10 keV) and the Large Area Detector (LAD; 2-30 keV) on board eXTP will allow us to accurately study the Fe-K region \citep[for more details of the onboard instrumentation, see][]{Zhang_19}. By simulating a 20 ks observation using {\tt Model 4} and removing the smeared reflection component, we should observe large residuals between 5 and 9 keV in the LAD spectrum (see Fig. \ref{figure:del_line}). Furthermore, the improved sensitivity of eXTP might allow us to discriminate between  {\tt Model 2} and {\tt Model 4}. 

 Finally, we also explored  the possibility that the residuals observed in the RGS12 spectrum could be fitted with 
a {\sc Bremss} component instead of a reflection component.  We found  that the bremsstrahlung temperature is only $0.31 \pm 0.03$ keV, while the electron temperature of the  Comptonized corona is close to 3 keV. This implies that the bremsstrahlung emission is produced far away from the central region of the system. From the normalization of the {\sc Bremss} component Norm$_{\rm br}$, we estimated the emission measure $n_e^2 V$ under the assumption that the matter is fully ionized and the ion density $n_I$ is equal to the electron density $n_e$. We obtained that $n_e^2V \simeq 2 \times 10^{59}$ cm$^3$.  Following the approach shown by \cite{Iaria_01b}, we can estimate the size of the emitting cloud assuming   an upper limit of the  optical depth $\tau$ of 1 and a spherical volume for the bremsstrahlung-emitting region. We obtained $R \gtrsim  2 \times 10^{10}$ cm and $n_e \lesssim 1.3 \times 10^{14}$ cm$^{-3}$; our result suggests that  an extended optically thin corona with a temperature of $0.31 \pm 0.03$ keV could be present. However, we note  that {\tt Model 2} and {\tt Model 4}, which contain the smeared reflection component, give  values of $\chi^2$ which are the lowest and the addition of the reflection component is more statistically significant than the {\sc Bremss} component.

 \section{Conclusions}
We presented the first 0.3-40 keV broadband analysis of the bright Atoll source GX 9+9.
%The main aspect is that although the spectrum does not a clear evidence of a broad emission line in the Fe-K region we show that a relativistic smeared reflection component is necessary to fit the broadband spectrum. 
The main result is the presence of a smeared relativistic reflection component, necessary to improve the fit of the broadband spectrum, although no clear evidence of a broad emission line in the Fe-K region is present.

The spectrum can be fitted with a disk-blackbody component plus a Comptonized component with seed photons having a blackbody spectrum associated with an optically thick boundary layer ({\tt Model 2}). An alternative best-fit model ({\tt Model 4}) is associated with the so-called  Western Model, in which we have a Comptonized component with seed photons having a disk-blackbody distribution plus a blackbody component that mimics a boundary layer or a transition layer \citep{Mainardi_10, Seifina_12}. As mentioned above, in both   cases a relativistic smeared reflection component is statistically required.  We estimated that the inclination angle is 
$43^{+6}_{-4}$ and $51^{+9}_{-2}$ degrees from  {\tt Model 2} and {\tt Model 4}, respectively. These values are compatible with each other 
within 90\% c.l. and compatible with the value provided by \cite{Hertz_88}, who suggested an inclination angle smaller than 63$^{\circ}$. The estimated unabsorbed 0.1-100 keV luminosity is $2.3 \times 10^{37}$ erg s$^{-1}$, assuming a distance to the source of 5 kpc. 

In {\tt Model 2}, the Comptonized component has a seed-photon temperature of $1.14^{+0.10}_{-0.07}$ keV and an electron temperature of $2.80^{+0.09}_{-0.04}$ keV; the estimated optical depth is  $8.9\pm0.4$. The color temperature of the innermost region of the accretion disk is  $0.86^{+0.08}_{-0.02}$ keV and $0.82 \pm 0.02$ keV from the SAX and XMM spectrum, respectively. The inner radius is $27 \pm 4$ km, compatible with the estimated value of the inner radius of the reflecting region ($21^{+4}_{-8}$). 

 In {\tt Model 4}, the Comptonized component has a seed-photon temperature of $0.87 \pm 0.07$ keV and $1.01 \pm 0.08$ keV for the SAX and XMM spectrum, respectively. The electron temperature is $2.9\pm 0.2$ keV; the estimated optical depth is $9.4^{+1.5}_{-1.1}$. The blackbody temperature associated with a saturated Comptonization is $1.79^{+0.09}_{-0.18}$ keV and $1.85^{+0.07}_{-0.15}$ keV  for the SAX and XMM spectrum, respectively.   The unabsorbed bolometric flux of the blackbody component is only 12\% of the total value.

\section*{Acknowledgements}

This research has made use of data and/or software provided by the High Energy Astrophysics Science Archive Research Center (HEASARC), which is a service of the Astrophysics Science Division at NASA/GSFC and the High Energy Astrophysics Division of the Smithsonian Astrophysical Observatory.\\
The authors acknowledge financial contribution from the agreement
ASI-INAF n.2017-14-H.0, from INAF mainstream (PI: T. Belloni; PI: A. De Rosa)
and from the HERMES project financed by the Italian Space
Agency (ASI) Agreement n. 2016/13 U.O.
RI and TDS acknowledge the research
grant iPeska (PI: Andrea Possenti) funded under the INAF
national call Prin-SKA/CTA approved with the Presidential Decree
70/2016. 

\bibliographystyle{aa}
\bibliography{biblio}
\end{document}